\documentclass[aps, prl, preprint, floatfix, superscriptaddress, onecolumn, showpacs]{revtex4-1}
\usepackage[usenames,dvipsnames,svgnames,table]{xcolor}
\definecolor{HanRed}{RGB}{208, 16, 76}
\usepackage{upgreek}
\usepackage{hyperref}
\hypersetup
{
    colorlinks=true,
    citecolor=blue,
    linkcolor=HanRed,
    filecolor=blue,
    urlcolor=blue
}
\usepackage{silence}
\usepackage{natbib}
\usepackage{tikz}
\usepackage{textcomp}
\usepackage{amsmath}
\usepackage{amssymb}
\usepackage{bm}

\usepackage[normalem]{ulem}
\usepackage{soul}
\usepackage{siunitx}
\usepackage[T1]{fontenc}

\usepackage{mathptmx}
\PassOptionsToPackage{unicode}{hyperref}
\PassOptionsToPackage{naturalnames}{hyperref}

\usepackage[pagewise]{lineno}

\begin{document}

\title{Enhancement in Thermally Generated Spin Voltage at Pd/NiFe$_\text{2}$O$_\text{4}$ Interfaces by the Growth on Lattice-Matched Substrates}

\author{A. Rastogi}
\thanks{These authors contributed equally to this work.}
\affiliation{Center for Materials for Information Technology, The University of Alabama, Tuscaloosa, Alabama 35487, USA}

\author{Z. Li}
\thanks{These authors contributed equally to this work.}
\affiliation{Center for Materials for Information Technology, The University of Alabama, Tuscaloosa, Alabama 35487, USA}
\affiliation{Department of Physics \& Astronomy, The University of Alabama, Tuscaloosa, Alabama 35487, USA}

\author{A. V. Singh}
\affiliation{Center for Materials for Information Technology, The University of Alabama, Tuscaloosa, Alabama 35487, USA}

\author{S. Regmi}
\affiliation{Center for Materials for Information Technology, The University of Alabama, Tuscaloosa, Alabama 35487, USA}
\affiliation{Department of Physics \& Astronomy, The University of Alabama, Tuscaloosa, Alabama 35487, USA}

\author{T. Peters}
\affiliation{Center for Spinelectronic Materials and Devices, Department of Physics, Bielefeld University, Universit\"atsstra\ss e 25, 33615 Bielefeld, Germany}

\author{P. Bougiatioti}
\affiliation{Center for Spinelectronic Materials and Devices, Department of Physics, Bielefeld University, Universit\"atsstra\ss e 25, 33615 Bielefeld, Germany}

\author{D. Carsten n\'e Meier}
\affiliation{Center for Spinelectronic Materials and Devices, Department of Physics, Bielefeld University, Universit\"atsstra\ss e 25, 33615 Bielefeld, Germany}

\author{J. B. Mohammadi}
\affiliation{Center for Materials for Information Technology, The University of Alabama, Tuscaloosa, Alabama 35487, USA}
\affiliation{Department of Physics \& Astronomy, The University of Alabama, Tuscaloosa, Alabama 35487, USA}

\author{B. Khodadadi}
\affiliation{Center for Materials for Information Technology, The University of Alabama, Tuscaloosa, Alabama 35487, USA}
\affiliation{Department of Physics \& Astronomy, The University of Alabama, Tuscaloosa, Alabama 35487, USA}

\author{T. Mewes}
\affiliation{Center for Materials for Information Technology, The University of Alabama, Tuscaloosa, Alabama 35487, USA}
\affiliation{Department of Physics \& Astronomy, The University of Alabama, Tuscaloosa, Alabama 35487, USA}

\author{R. Mishra}
\affiliation{Department of Mechanical Engineering and Materials Science, and Institute of Materials Science and Engineering, Washington University in St. Louis, St. Louis, Missouri 63130, USA}

\author{J. Gazquez}
\affiliation{Institut de Ci\`{e}ncia de Materials de Barcelona, Campus de la UAB, 08193, Bellaterra, Spain}

\author{A. Y. Borisevich}
\affiliation{Materials Science and Technology Division, Oak Ridge National Laboratory, TN 37831, USA}

\author{Z. Galazka}
\affiliation{Leibniz-Institut f\"ur Kristallz\"uchtung, Max-Born-Str.\,2, 12489 Berlin, Germany}

\author{R. Uecker}
\affiliation{Leibniz-Institut f\"ur Kristallz\"uchtung, Max-Born-Str.\,2, 12489 Berlin, Germany}

\author{G. Reiss}
\affiliation{Center for Spinelectronic Materials and Devices, Department of Physics, Bielefeld University, Universit\"atsstra\ss e 25, 33615 Bielefeld, Germany}

\author{T. Kuschel}
\email[E-mail: ]{tkuschel@physik.uni-bielefeld.de}
\affiliation{Center for Spinelectronic Materials and Devices, Department of Physics, Bielefeld University, Universit\"atsstra\ss e 25, 33615 Bielefeld, Germany}

\author{A. Gupta}
\email[E-mail: ]{agupta@mint.ua.edu}
\affiliation{Center for Materials for Information Technology, The University of Alabama, Tuscaloosa, Alabama 35487, USA}

\date{\today}

\begin{abstract}

Efficient spin injection from epitaxial ferrimagnetic NiFe$_2$O$_4$ thin films into a Pd layer is demonstrated via spin Seebeck effect measurements in the longitudinal geometry. The NiFe$_2$O$_4$ films (60 nm to 1 \si{\um}) are grown by pulsed laser deposition on isostructural spinel MgAl$_2$O$_4$, MgGa$_2$O$_4$, and CoGa$_2$O$_4$ substrates with lattice mismatch varying between 3.2\% and 0.2\%. For the thinner films ($\le$ 330 nm), an increase in the spin Seebeck voltage is observed with decreasing lattice mismatch, which correlates well with a decrease in the Gilbert damping parameter as determined from ferromagnetic resonance measurements. High resolution transmission electron microscopy studies indicate substantial decrease of antiphase boundary and interface defects that cause strain-relaxation, i.e., misfit dislocations, in the films with decreasing lattice mismatch. This highlights the importance of reducing structural defects in spinel ferrites for efficient spin injection. It is further shown that angle-dependent spin Seebeck effect measurements provide a qualitative method to probe for in-plane magnetic anisotropies present in the films.

\end{abstract}
\maketitle

\section{I. INTRODUCTION}

Efficient conversion of heat to electric energy in thermo-electric materials is an active field of research. Recent studies on the interaction between electron spin and heat flow have created a new area of research in spintronics that is commonly referred to as spin caloritronics~\cite{wolf2001spintronics,vzutic2004spintronics,Bauer2012,Kirihara2012,Uchida2011,Boona2014a,uchida2016}. The spin Seebeck effect (SSE), which involves generation of spin current through heat flow, is one of the most promising phenomena in the emerging field of spin caloritronics. One approach to efficiently generate spin current is the implementation of a temperature gradient across a magnetic thin film that is perpendicular to the magnetization~\cite{Uchida2008,Bosu2011,Jaworski2010,Adachi2010,Meier2013a}. The spin current is generated parallel to the temperature gradient via the so-called longitudinal spin Seebeck effect (LSSE). It can be injected into a normal metal (Pt, Pd, Au, etc.) electrode and converted into a charge current due to the inverse spin Hall effect (ISHE)~\cite{valenzuela2006so,saitoh2006,kimura2007room}. The electric field ($\bm{E}_{\text{ISHE}}$) generated by the spin current in a normal metal is described by the relationship~\cite{Adachi2010}

\begin{equation}\label{Eq_SSE}
    \bm{E}_{\text{ISHE}}=\theta_{\text{SH}} \, \rho \, \bm{J}_{\text{s}} \times \bm{\sigma}\, ,
\end{equation}
where $\theta_{\text{SH}}$ is the spin-Hall angle, $\rho$ is the electrical resistivity of the normal metal, $\bm{J}_{\text{s}}$ is spin current density, and $\bm{\sigma}$ is spin-polarization vector, collinear with the magnetization $\bm{M}$.

Using magnetic insulators as a source of spin current has advantages over magnetic metals because unintended effects such as the anomalous Nernst effect can be neglected due to the absence of conduction electrons~\cite{Huang2011}. In magnetic insulators, magnons, the quanta of spin waves, are the carriers of the generated spin current.

Yttrium iron garnet (YIG) is the most widely studied insulating ferrimagnetic material for LSSE experiments because of its low magnetic coercivity and an extremely low Gilbert damping~\cite{hauser2016yttrium}. Nickel ferrite (NiFe$_2$O$_4$, NFO) is also a promising candidate for high frequency applications as its saturation magnetization is much higher than YIG~\cite{chinnasamy2007effect}. The use of NFO has further advantages such as the tuning of electrical properties by temperature~\cite{Meier2013a} or by oxygen content~\cite{Bougiatioti2017,bougiatioti2017electrical}. However, so far there have been only few reports of LSSE using NFO thin films. The NFO films used in previous studies were deposited by either chemical vapor deposition method~\cite{Meier2013a,Meier2015,meier2016detection,Kuschel2015} or reactive co-sputtering~\cite{klewe2014physical,Kuschel2016,Shan2017a,Bougiatioti2017} on MgAl$_2$O$_4$ substrate that has a large lattice mismatch of $\sim$3.2\%, resulting in the formation of antiphase boundaries (APBs) and interface defects, such as misfit dislocations~\cite{li2012microstructural}, which limits their usability for device applications. Nevertheless, recent nonlocal magnon spin transport experiments~\cite{Shan2017a} based on the SSE in sputter-deposited NFO on MgAl$_2$O$_4$ show that the magnon spin diffusion length is $\sim$3 \si{\um}, which is in the same range as for YIG~\cite{cornelissen2015long}. We have recently shown that with appropriate choice of substrates and growth conditions, NFO thin films can exhibit a saturation magnetization as high as its bulk value, with damping constant and coercivity values comparable to that of YIG~\cite{Singh2017}. Moreover, Pd is another metal with high spin Hall angle besides Pt, which shows strong potential for spintronics applications~\cite{ando2010inverse,tao2018self,ma2018spin}.

In this work, we report on a systematic study of enhancement in the thermally generated ISHE voltage for Pd/NFO films on different (001)-oriented isostructural spinel substrates: MgAl$_2$O$_4$ (MAO), MgGa$_2$O$_4$ (MGO), and CoGa$_2$O$_4$ (CGO) with decreasing lattice mismatch of $\sim$3.2\%, 0.8\%, and 0.2\% with NFO, respectively. The overall microstructure and the interface between the films and substrates have been investigated by high resolution scanning transmission electron microscopy (STEM), which shows a substantial decrease of APBs and misfit dislocations with decreasing lattice mismatch. For thinner films ($\leq$ 330 nm), the obtained LSSE results correlate well with the damping parameters as determined by ferromagnetic resonance measurements (FMR). The thermally generated spin voltage signal increases with decreasing lattice mismatch, whereas the damping parameter decreases.

\section{II. Experimental}

\noindent \textbf{1. Sample preparation and characterization}\\

High-quality epitaxial NFO thin films were deposited using pulsed laser deposition followed by in-situ Pd deposition by DC sputtering. For NFO film deposition we used a laser fluence of $\sim$1 J/cm$^2$ in an oxygen environment with a background pressure of 1.3 Pa. The temperature of the substrates was kept constant at 700 $^{\circ}$C during film growth. We used three different (001)-oriented spinel substrates, namely MAO, MGO, and CGO. The MAO substrates were purchased commercially (CrysTec GmbH), while the MGO and CGO substrates were prepared from high quality single crystals, which were grown at the Leibniz Institute for Crystal Growth~\cite{galazka2015} and then cut and polished by CrysTec GmbH, Berlin, Germany. We investigated films with thicknesses ranging from 60 nm to 1 \si{\um} deposited on substrates with a size of 3$\times$5 mm$^2$. For LSSE measurements, the deposition of NFO film was followed by in-situ deposition of a 5 nm thick Pd layer by DC sputtering at 0.7 Pa Argon pressure and 20 W power. 

The films were structurally characterized using a Philips X$'$Pert X-ray diffractometer. High resolution STEM and high-angle annular dark field imaging (along the [001] direction) were carried out on some of the samples in an aberration-corrected Nion UltraSTEM$^{\text{TM}}$ 200 microscope operating at 200 kV. Two different imaging modes were used, the high-angle annular dark field (HAADF) and the low-angle annular dark field (LAADF) imaging modes. The HAADF imaging mode gives rise to the so-called Z-contrast, and it was acquired using an annular detector with a high inner collection angle~\cite{pennycook1991high}. On the other hand, the LAADF imaging mode is achieved using an annular detector with a smaller inner collection angle, which allows collection of electrons scattered by the strained regions giving rise to different angular distributions of the annular dark field signal, thus causing extra contrast~\cite{cowley1992channelling}.

The films were magnetically characterized using vibrating sample magnetometry (VSM) in a PPMS$^\circledR$ DynaCool$^\text{TM}$ system (Quantum Design). Room temperature broadband ferromagnetic resonance (FMR) measurements were performed using a coplanar waveguide to determine the effective Gilbert damping parameter of two films deposited on MGO and CGO substrates. The FMR measurements were carried out in the in-plane geometry, i.e. with the quasi-static magnetic field applied in the plane of the film. \\

\noindent \textbf{2. Measurement setup for spin Seebeck effect}\\

\begin{figure}[!htb]
    \centering
    \includegraphics[width=\linewidth]{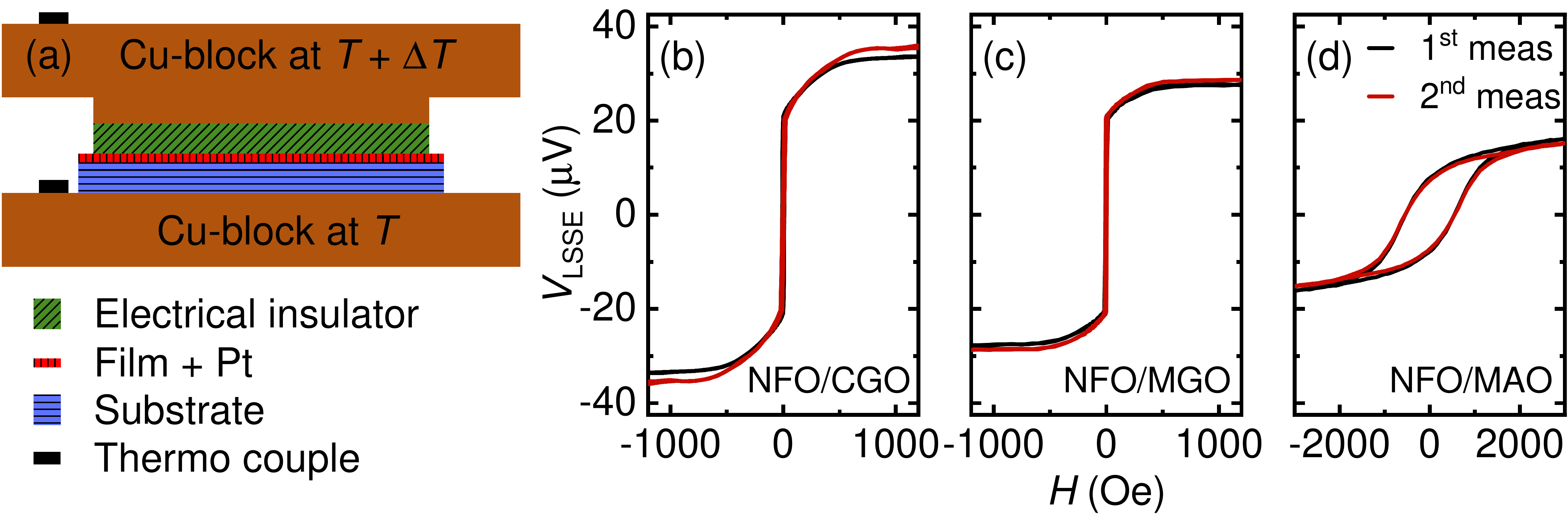}
    \caption{(a) A schematic of the measurement setup for temperature gradient method. Panel (b), (c), and (d) show the reproducibility of the $V_{\text{ISHE}}$ signal using SiC spacer. The measurements were done on 330 nm thick NFO films deposited on CGO, MGO, and MAO. Black lines show the results of first measurement, while the red lines show a repeat measurement after remounting the same sample.}
    \label{FIG1_Schematic}
\end{figure}

We used two methods to normalize the $V_{\text{ISHE}}$ signal, namely by heat flux and by thermal gradient. For the heat flux setup in Bielefeld, we used a calibrated Peltier element clamped between the sample and one of the copper blocks to detect the heat flux as described in Ref.~\cite{sola15,Sola2017,sola2018,Bougiatioti2017}. The heat flux method developed by Sola \textit{et al.} helps to improve the reproducibility when determined LSSE coefficients are compared between different setups as well as when remounting samples in the same setup~\cite{Sola2017,sola2018}. In the thermal gradient setups in Alabama and Bielefeld, only the sample was sandwiched between two copper blocks (Fig.~\ref{FIG1_Schematic}(a)). The Cu-blocks were retained in good thermal contact with Peltier elements for cooling and heating. A thermally conducting and electrically insulating 250 \si{\um} thick SiC spacer was used between the top Pd layer and the upper copper block. For a comparison of different spacers, see Fig.~\textcolor{HanRed}{S2} in \textcolor{HanRed}{SI}. For all measurements the spacing between the voltage probes ($w$) was kept constant, $w$ $\approx$ 4.8 mm. The temperature of the lower block was fixed at a base temperature $T$ (room temperature, if not stated otherwise), while the temperature of the upper block was varied ($T + \Delta T$) to obtain the desired temperature difference across the sample. A K-type thermocouple was used to measure the temperature at each Cu-block. For angular-dependent measurements the sample was rotated in-plane with a manual stage. A helium-based closed cycle refrigerator was used to carry out the low-temperature measurements. To check the reproducibility of the voltage signal in our setup, we have remeasured the same sample repeatedly after remounting, but the voltage signal remains unaffected within the error limit as shown in Fig.~\ref{FIG1_Schematic}(b), \ref{FIG1_Schematic}(c) and \ref{FIG1_Schematic}(d). The primary source of error in our measurements is the distance between the electrical contact ($\sim$4\%). Since the voltage signal remains essentially unchanged after repeated measurements, we can compare results from the same setup using the temperature difference method in addition to the heat flux technique. We used the temperature gradient method for the LSSE measurements of magnetic field and temperature variations. For a quantitative comparison of substrate effects in LSSE, we used the heat flux method.

\section{III. Results and discussion}

\noindent \textbf{A. Structural characterization of NFO films on different substrates}\\

\begin{figure}[!htb]
    \centering
    \includegraphics[width=\linewidth]{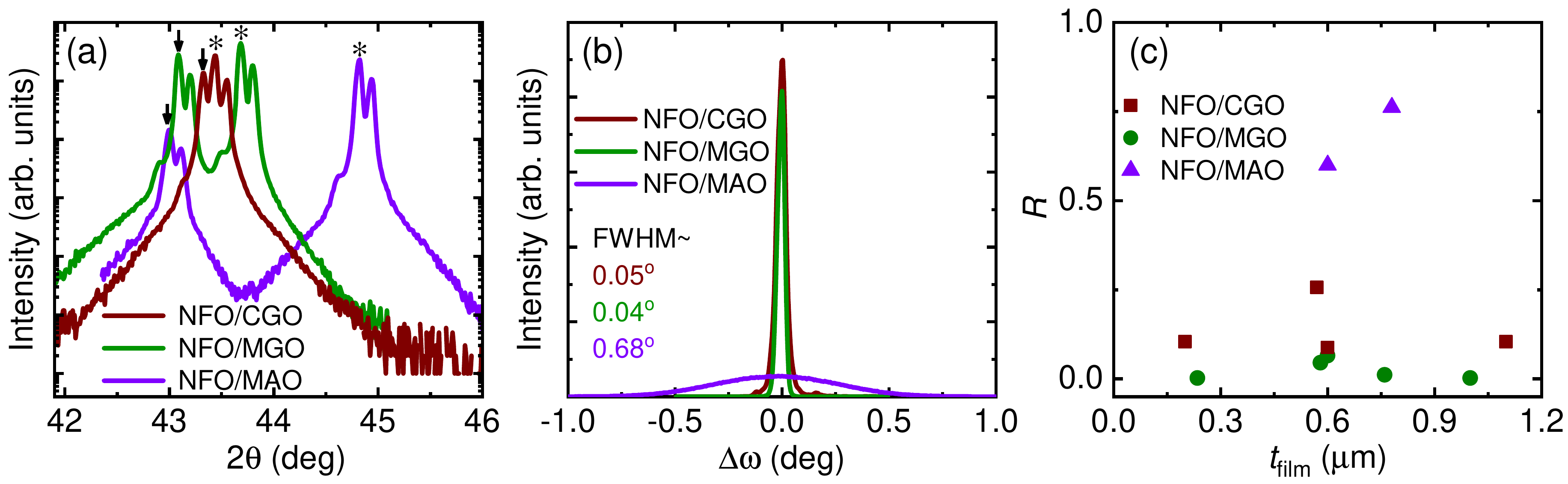}
    \caption{(a) Standard $\theta$-2$\theta$ diffraction patterns around the (004) reflections of the substrates (*) and films ($\downarrow$), respectively. (b) The full width at half maximum (FWHM) of omega scans of NFO films grown on the different substrates. (c) Variation of the strain parameter $R$ with the thickness of NFO films deposited on the different substrates. The films on MAO substrate are closer to being relaxed.}
    \label{FIG2_XRD}
\end{figure}

All three (001)-oriented substrates, namely CGO, MGO, and MAO, impose a compressive strain on the NFO film, and hence the lattice parameter elongates in the out-of-plane direction. It can be seen in Fig.~\ref{FIG2_XRD}(a) that the film peak position shifts to lower values of 2$\theta$ (2$\theta_{\text{bulk}} = 43.33^{\circ}$) with increasing lattice mismatch. Omega scans in Fig.~\ref{FIG2_XRD}(b) indicate that epitaxial quality of the films on MGO and CGO are significantly better than the film on MAO. We also performed off-axis XRD scans on few films deposited on three different substrates and calculated the strain in the films which can be quantified by the parameter $R=(a_\text{f}-a_\text{s})/(a_\text{b}-a_\text{s}$), with a$_\text{f}$, a$_\text{b}$, and a$_\text{s}$ as in-plane lattice parameters of the NFO thin film (measured), NFO bulk (literature), and substrate (single crystal), respectively. Therefore, $R = 1$ for a fully relaxed film and $R = 0$ for a fully strained film. As shown in Fig.~\ref{FIG2_XRD}(c), the films on CGO and MGO substrates have significantly lower values of R than those on MAO and are not fully relaxed with even the thickest films remaining strained. We also used X-ray reflectivity technique to determine the Pd layer thickness, which is essentially the same ($\sim5.0 \pm 0.4$ nm) for all the samples.

\begin{figure}[!htb]
\centering
\includegraphics [width=\linewidth]{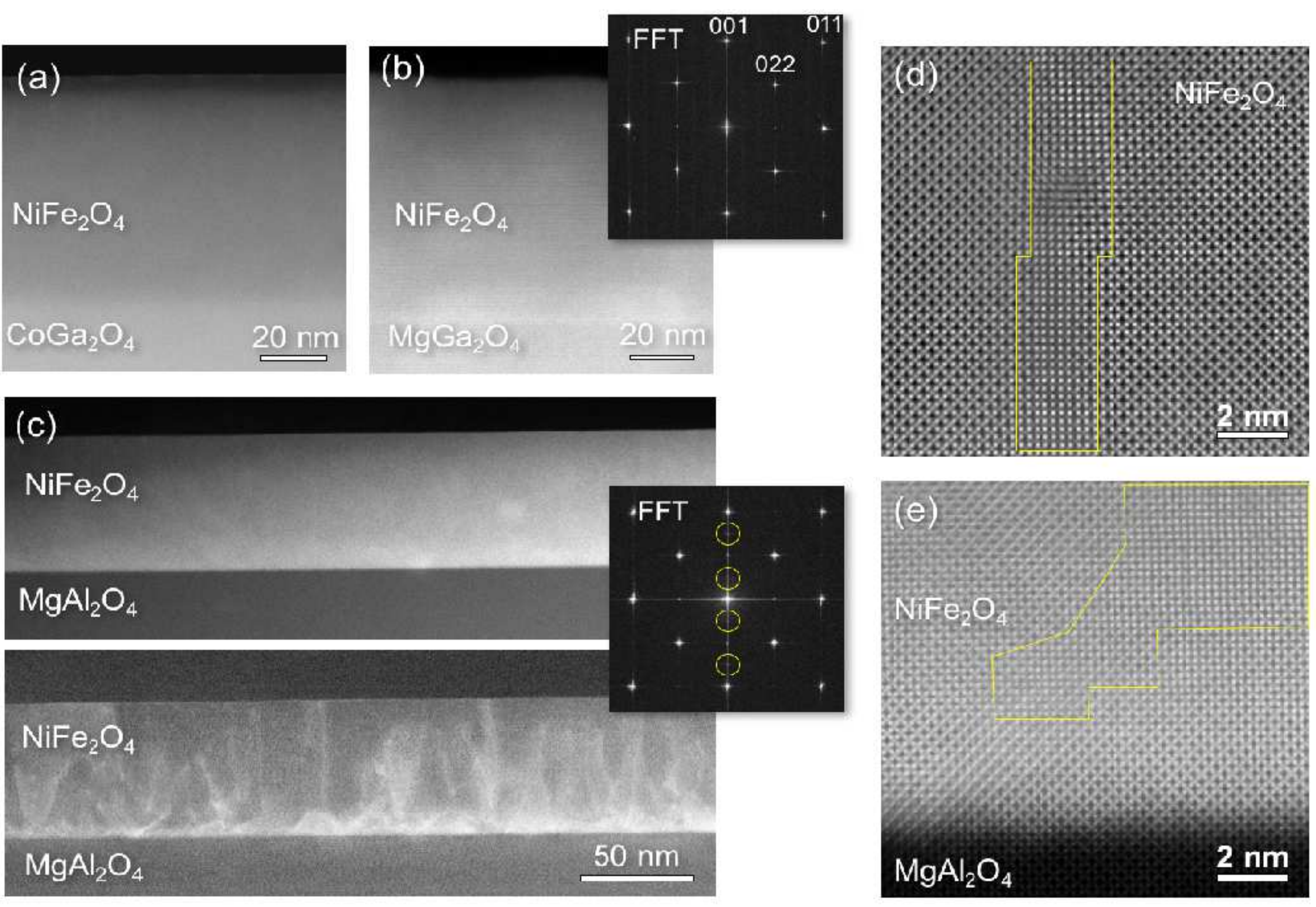}
\caption {(a) and (b) show low magnification HAADF Z-contrast images of NFO films ($\sim$60 nm) grown on CGO and MGO substrates, respectively. The inset shows a characteristic FFT pattern from the Z-contrast image of (b). (c) Upper and lower panels show low magnification Z-contrast and LAADF images of NFO films grown on MAO, respectively. The inset shows an FFT of the NFO film grown on MAO substrate. The yellow circles highlight the extra reflections arising due to the APBs. High resolution Z-contrast images of an APB within the bulk of the film (d) and close to the interface (e) of the NFO films grown on MAO. APBs are highlighted in yellow.}
\label{FIG3_STEM} 
\end{figure}

Low magnification STEM images of the films grown on CGO, MGO and MAO substrates are shown in Fig.~\ref{FIG3_STEM}(a), \ref{FIG3_STEM}(b) and \ref{FIG3_STEM}(c), respectively. Two different imaging modes are used, the high-angle annular dark field (HAADF) and the low-angle annular dark field (LAADF) imaging modes. While the films grown on CGO and MGO exhibit sharp interfaces and are essentially free of APBs and other defects, the film grown on MAO presents many structural defects. These defects are clearly seen using the LAADF imaging mode, as shown in the lower panel of Fig.~\ref{FIG3_STEM}(c). The bright contrast of this image stems from crystal defects, mainly APBs, with a crystallographic translation of 1/4a [001]. In high resolution STEM Z-contrast images they appear with a clear distinct contrast, as highlighted in Figs.~\ref{FIG3_STEM}(d) and \ref{FIG3_STEM}(e). These defects also appear as a superstructure in fast Fourier transform (FFT) patterns, as shown in the FFT of an NFO film grown on MAO substrate (inset Fig.~\ref{FIG3_STEM}(c)). The extra reflections marked with yellow circles in the FFT are due to the presence of APBs, and are absent in the FFT patterns of NFO films grown on CGO and MGO substrates (inset Fig.~\ref{FIG3_STEM}(b)). The LAADF image also shows that the defects are unevenly distributed, as the density of APBs decreases near the surface of the film. Our previous studies have established that even relatively thick NFO films (100-450 nm) grown on CGO and MGO substrates remain essentially fully strained while those on MAO are partially relaxed with formation of misfit dislocations~\cite{Singh2017}. This is consistent with the X-ray diffraction results. The films on MAO also show presence of threading dislocation and dark diffused contrast areas, likely from A-site cation vacancies~\cite{li2012microstructural,Singh2017}. The APBs and other structural defects are known to cause a reduction of saturation magnetization and increase in the FMR linewidth of the thin films compared to their bulk values~\cite{torres1993effect}. However, their effect on the spin transport properties and especially on ISHE remain unknown.\\

\noindent \textbf{B. Spin Seebeck effect measurements of NFO films on different substrates}\\

\begin{figure*}[!htb]
\centering
\includegraphics [width=\linewidth]{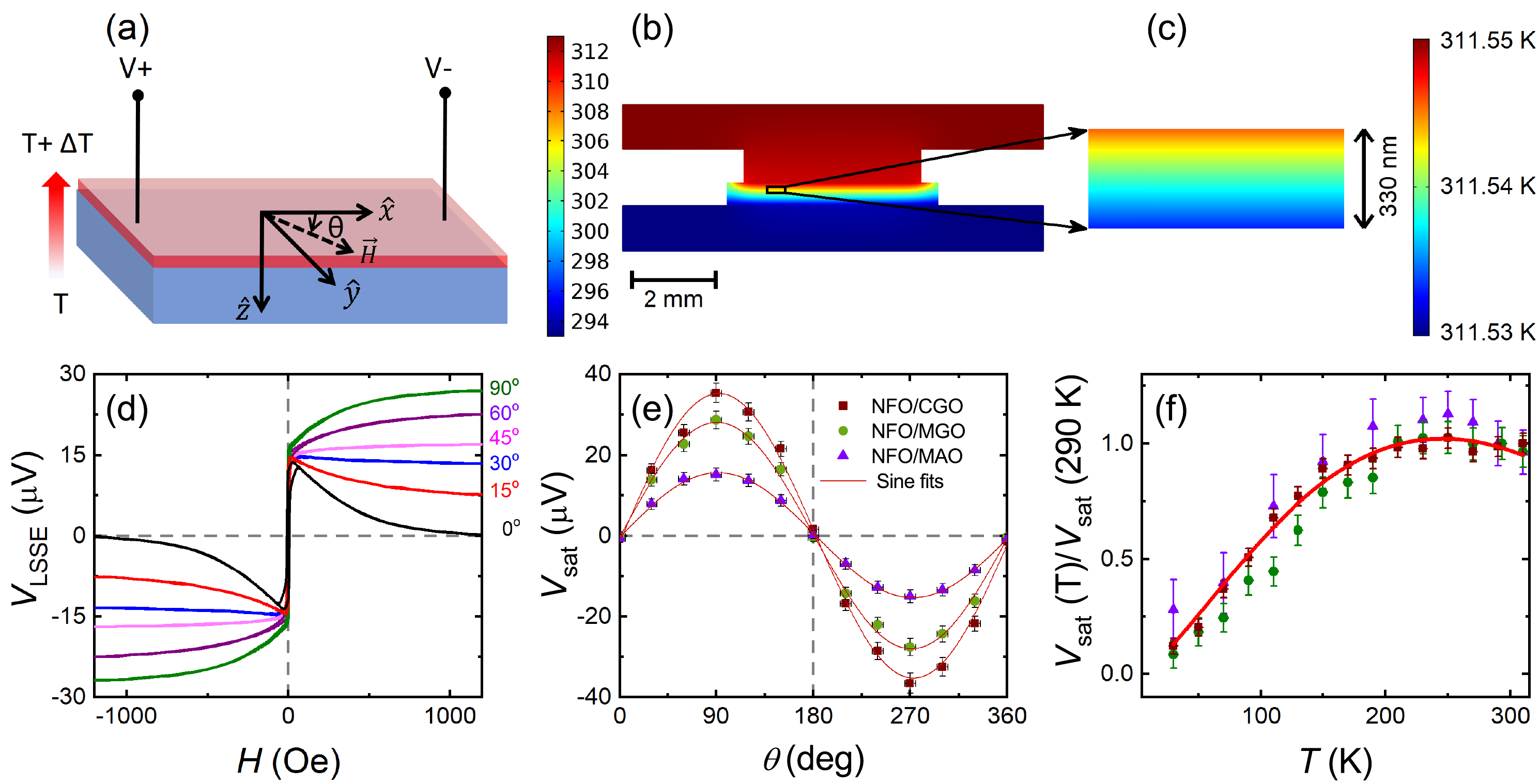} 
\caption{(a) A schematic of the LSSE measurement geometry. A temperature gradient is created along the $\hat{z}$-direction; the magnetic field is applied in the sample plane with an angle $\theta$ with respect to the $\hat{x}$-direction, and the voltage is measured in the same plane. (b) and (c) Results of COMSOL Multiphysics$^\circledR$ simulation for the generation of the temperature gradient across the sample and the heating components. The film thickness is $\sim$330 nm. (d) LSSE measurements for Pd/NFO/MGO (001) with voltage contacts located along the $\hat{x}$-direction and the external magnetic field applied in-plane at various angles with respect to the voltage contacts. A complete angular dependence of the saturation voltage for all the three films (330 nm) is plotted in panel (e); the dotted lines are sine function fits. (f) Variation of normalized voltage signal ($V_{\text{sat}}(T)/V_{\text{sat}}$(290K)) with various base temperature ($T$) for 330 nm thick films on different substrates. The solid line is fit to the NFO/CGO data using the Eq.~\ref{Eq_SSE} described in the text.} 
\label{FIG4_COMSOL_SSE} 
\end{figure*}

In Fig.~\ref{FIG4_COMSOL_SSE}(a), we show a schematic of the measurement geometry for LSSE. The temperature gradient across the film and the substrate has been simulated using the heat transfer module and finite element method available in COMSOL Multiphysics$^\circledR$. The simulation for a 330 nm NFO film on MGO substrate is shown in Fig.~\ref{FIG4_COMSOL_SSE}(b) and \ref{FIG4_COMSOL_SSE}(c). The temperature gradient ($\Delta T_\text{f}$) is in the range of tens of mK/\si{\um} when a temperature difference of $\sim$20 K is applied across the Cu-blocks. Fig.~\ref{FIG4_COMSOL_SSE}(b) shows the cross-sectional view of temperature distribution across the stack. For clarity, the temperature profile across the film is shown in the zoomed cross-section image (Fig.~\ref{FIG4_COMSOL_SSE}(c)). Further details are provided in the Supplementary Information (\textcolor{HanRed}{SI}) section \textcolor{HanRed}{I}. We find that the temperature difference across the film scales with the temperature difference across the Cu-blocks and is essentially independent of the choice of the substrate (MAO, MGO and CGO) because of their similar thermal characteristics (see Table~\textcolor{HanRed}{I} in \textcolor{HanRed}{SI}). 

In our geometry we are sensitive to the $\hat{x}$-component of $\bm{E}_{\text{ISHE}}$ (with $V_{\text{ISHE}} = {E}_{\text{ISHE}}\cdot{w}$, $w$ is the distance between voltage probes), and according to Eq.~\ref{Eq_SSE} we are sensitive to the $\hat{y}$-component of $\bm{\sigma}$ and thus $\bm{M}$. The background signal is subtracted from data presented here (for raw data please see Fig.~\textcolor{HanRed}{S3} in \textcolor{HanRed}{SI}). In Fig.~\ref{FIG4_COMSOL_SSE}(d), we display the result for a 330 nm thick NFO/MGO film with angular variation from 0$^\circ$ to 90$^\circ$ between the voltage contacts and the magnetic field. We observe that upon reversing the direction of $\Delta T_\text{z}$, the voltage signal is also reversed, which is a characteristic behavior of $V_{\text{ISHE}}$ induced by LSSE (see Fig.~\textcolor{HanRed}{S4} in \textcolor{HanRed}{SI}). To obtain the maximum LSSE voltage the external magnetic field is applied along the $\hat{y}$-direction ($\theta$ = 90$^\circ$) to saturate the magnetization aligned along this direction. This leads to a maximum $V_{\text{sat}}$ of about $\sim$27 $\upmu$V. After magnetic field reversal the magnetization direction is changed into the opposite direction and $V_\text{sat}$ of $\sim$-27 $\upmu$V is obtained. During the magnetic field reversal process (Fig.~\ref{FIG4_COMSOL_SSE}(d)), $V_\text{LSSE}$ acts in correspondence with the magnetization and correlates well with the VSM measurement (see Fig.~\textcolor{HanRed}{S5} in \textcolor{HanRed}{SI}). When $\theta$ is reduced, $V_\text{sat}$ decreases and follows the cross product of Eq.~\ref{Eq_SSE}, which is evident from Fig.~\ref{FIG4_COMSOL_SSE}(e). During the magnetic field reversal process, the magnetization rotates towards one of the magnetic easy axes aligned along 45$^\circ$ in [011] directions~\cite{pachauri2016comprehensive}. For angles $\theta > 45^\circ$ upon reducing the magnetic field, the projection of the magnetization onto the $\hat{y}$-direction also decreases which results in a decrease of $V_\text{LSSE}$. For $\theta = 45^\circ$, $V_\text{LSSE}$ signal shows the maximum squareness while the magnetization lies along one of the magnetic easy axes. For angles $\theta < 45^\circ$, $V_\text{LSSE}$ signal increases when the magnetic field is decreased due to the increase of the projection of the magnetization in the $\hat{y}$-direction. Across $H$ = 0 Oe, $\bm{M}$ lies along one of the magnetic easy axes and results in nearly the same remanent voltage signal for all angles $\theta$ (see Fig.~\textcolor{HanRed}{S6} in \textcolor{HanRed}{SI}). For $\theta$ $\le$ 30$^\circ$ we observe a slight difference around $H$ = 0 Oe, which can be attributed due to the multi domain formation during the reversal process~\cite{Kehlberger2014}. We have additionally performed magnetic and LSSE measurements on an NFO film grown on (011)-oriented MGO substrate (see Section III in \textcolor{HanRed}{SI} or the results in Ref.~\cite{li2019}).

The temperature dependence (from 30 K to 300 K) of normalized LSSE voltage for 330 nm thick films is shown in Fig.~\ref{FIG4_COMSOL_SSE}(f), with the $\Delta$T across the stack being fixed at 20 K. This observation is similar to the results for CVD deposited NFO films on MAO substrate~\cite{Meier2013a}. In some previous reports the temperature dependence of the ISHE signal has been discussed for magnetic insulator/normal metal hybrid structures~\cite{Weiler2013,Schreier2013a,Uchida2014c}. A $T^{3/2}$ variation at low temperatures has been theoretically proposed,~\cite{Weiler2013,Schreier2013a} while a $(T_\text{c}-T)^{3}$ ($T_\text{c}$ is the Curie temperature) dependence at higher temperatures has been experimentally observed for Pt/YIG~\cite{Uchida2014c}. We combined these two temperature regimes and fitted our data (Fig.~\ref{FIG4_COMSOL_SSE}(f)) with $V_\text{LSSE}\propto T^{3/2} (T_\text{c}-T)^3$. This relationship fits well with our observation in the measured temperature range. From the fits, the $T_\text{c}$ is found to be in the range 700 K -- 800 K, which is close to NFO bulk value ($\sim$850 K)~\cite{luders2006}.

\begin{figure}[!htb]
\centering
\includegraphics[width=0.8\linewidth]{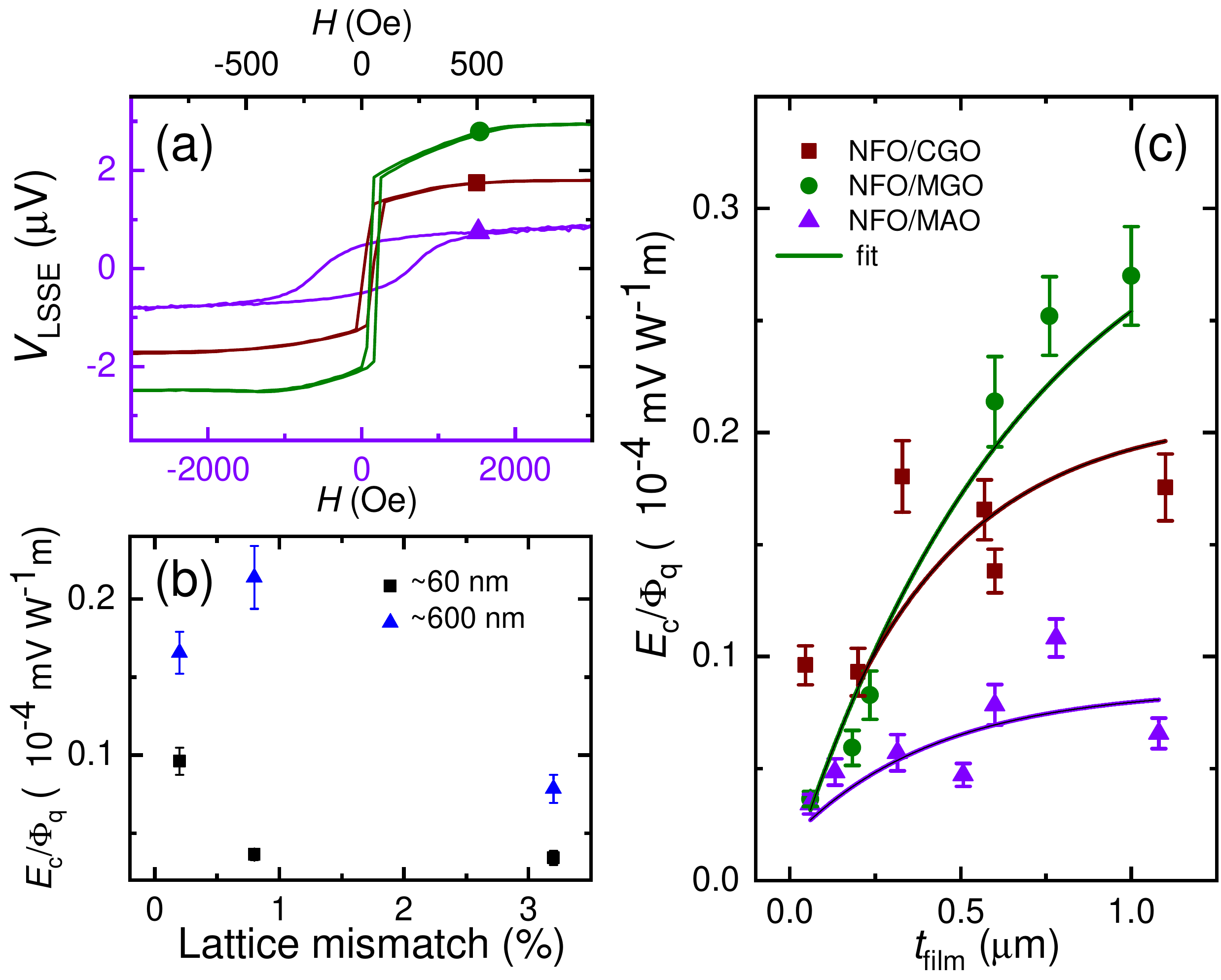}
\caption{(a) Magnetic field dependence of spin voltage signal generated at constant heat flux ($\sim$24 kW/m$^2$) for three Pd/NFO films (600 nm) on different substrates. (b) Influence of lattice mismatch of NFO film with CGO, MGO, and MAO substrates for two different film thicknesses. (c) Thickness dependence for films on different substrates (symbols), while solid lines are fit to the equation $E_{\text{LSSE}} \propto 1-\text{exp}(-t_\text{{film}}/\xi$), $\xi$ is the magnon propagation length~\cite{Kehlberger2015}. All the measurements were performed using heat flux method at room temperature.}
\label{FIG5_SSE_heat_flux} 
\end{figure}

In Figure~\ref{FIG5_SSE_heat_flux}(a), we plot the magnetic field variation of the SSE voltage of 600 nm thick films on MGO (circle), CGO (square), and MAO (triangle) obtained using the heat flux method. Here the $V_\text{LSSE}$ signal of the film on MGO is larger as compared to CGO. On the other hand, SSE voltage across the films on MAO substrate remains lowest in both the measurement techniques which is evident from Fig.~\ref{FIG4_COMSOL_SSE}(e) and Fig.~\ref{FIG5_SSE_heat_flux}(a). In Fig.~\ref{FIG5_SSE_heat_flux}(b), we show the variation of normalized saturation electric field ($E_\text{c}$) generated in the Pd-layer as a function of the lattice mismatch with the three substrates. We observe a weak SSE signal for films grown on MAO substrate and larger SSE response for NFO films on MGO and CGO. Overall, it is noted that irrespective of the thickness of the films, MAO substrate shows the lowest LSSE signal. This signifies the importance of lattice mismatch in enhancing the $\frac{E_\text{c}}{\Phi_\text{q}}$ signal. In conjunction with the STEM results we conclude that APBs and other structural defects present in the films are one of the reasons associated with the change in the LSSE signal. The values of  $\frac{E_\text{c}}{\Phi_\text{q}}$ for Pd/NFO/MGO are in a similar range ($\sim$30 nm/A) as recently reported for Pt/YIG/GGG thin film heterostructure ($\sim$40 nm/A)~\cite{Prakash2018}. Here, the effect of the lower spin Hall angle of Pd~\cite{ando2010inverse,tao2018self,ma2018spin} is probably compensated by a larger SSE in the NFO. If directly compared to sputter-deposited Pt/NFO bilayers ($\sim$100 nm/A)~\cite{Bougiatioti2017}, the effect of less efficient spin-to-charge conversion in Pd becomes obvious. However, complete SSE thickness dependencies are quite rare in the literature, especially when normalized to the heat flux, and should be investigated in future studies.
Finally, since the spin Seebeck resistivity $\frac{E_\text{c}}{\Phi_\text{q}}$ is only an effective SSE coefficient that still includes the heat conductivity of the NFO, any thickness-dependent change of the NFO heat conductivity can affect the thickness-dependence of the heat-flux-normalized SSE voltages. The study of this dependence will be part of future work.

Our measurements show an increase in normalized saturation electric field ($E$) generated across the Pd-layer by the heat flux ($\Phi_{\text{q}}$) with increasing film thickness (Fig.~\ref{FIG5_SSE_heat_flux}(c)), which can be explained based on characteristic magnon propagation length ($\xi$), i.e. the number of magnons reaching the Pd/ferrimagnetic interface increases with thickness~\cite{ritzmann2014propagation,Prakash2018,Kehlberger2015} and contributes to the voltage signal. The value of $\xi$ deduced from the fits  is in the range of 400 to 700 nm. This is lower than the recently reported value obtained from nonlocal magnon spin transport measurements in sputter deposited NFO films ($\sim$3 \si{\um})~\cite{Shan2017a,shan2018}. Such discrepancy between local LSSE and nonlocal magnon spin transport results has also been observed for YIG~\cite{Kehlberger2015,cornelissen2015long} and can be explained by the different nature of the experiments. While magnons with different propagation lengths can reach the Pd interface in the local experiment, the magnons with small diffusion length cannot make it to the Pd detector in the nonlocal geometry. Upon further increasing the film thickness, we observe an increase/saturation in the voltage signal. Significant scatter is observed in the data points for the films on CGO. This might be due to differences in the quality of CGO substrates which is also reflected in FMR measurements, where we find scatter in the FMR linewidth (see Fig.~\textcolor{HanRed}{S8} in \textcolor{HanRed}{SI}).\\

\noindent \textbf{C. FMR measurements of NFO films on different substrates}\\

\begin{figure}[!htb]
\centering
\includegraphics[width=0.8\linewidth]{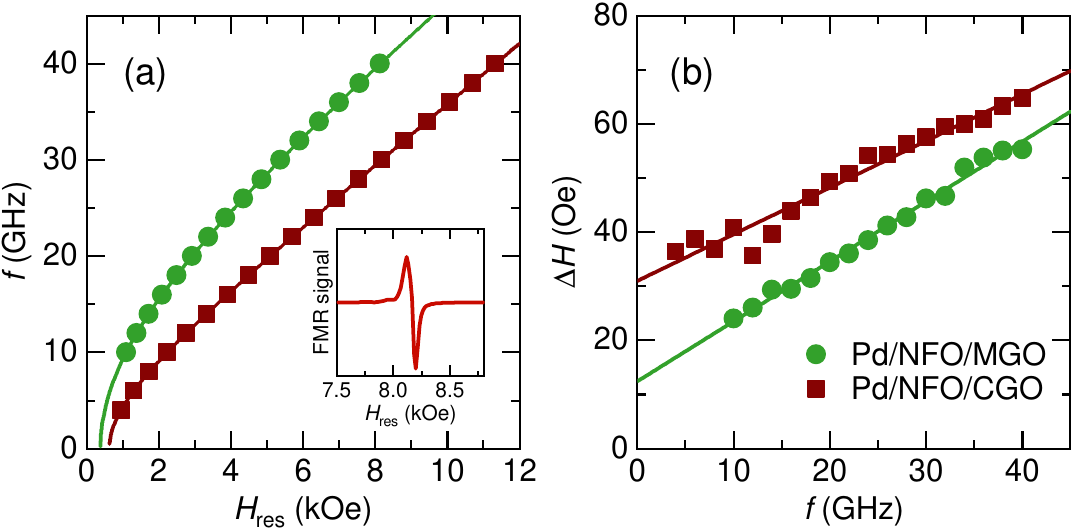}
\caption{Broadband FMR measurement results of 330 nm thick NFO films with Pd top layer deposited on CGO and MGO substrates. (a) Microwave frequency vs.\ resonance field data (symbols) are fitted to Kittel's equation (solid lines). The inset shows a typical FMR spectra of the film at 30 GHz frequency. (b) The dependence of FMR line width signal with resonance frequency (solid data points) and solid lines are fits to calculate the effective Gilbert damping and inhomogeneous linewidth broadening.}
\label{FIG6_FMR} 
\end{figure}

In addition to the LSSE measurements, we compared the dynamical properties of NFO films (330 nm) deposited on CGO and MGO substrates by FMR in the field along the in-plane hard axis geometry. From the measurements, we have estimated the effective magnetization ($M_\text{eff}$) and gyromagnetic ratio ($\gamma'$) from fitting the frequency ($f$) versus resonance field (H$_\text{res}$) data (Fig.~\ref{FIG6_FMR}(a)) to the Kittel equation in the in-plane configuration using equation $f=\gamma'\sqrt{(H_\text{res}+H_\text{4})\cdot(H_\text{res}+H_\text{4}+4\pi M_\text{eff})}$ with $H_\text{4}$ being the four-fold in-plane anisotropy. The FMR linewidth ($\Delta H$) vs.\ frequency ($f$) data is then used to calculate the effective Gilbert damping parameter ($\alpha_\text{{eff}}$) and inhomogeneous linewidth broadening ($\Delta H_0$) from $\Delta H = \Delta H_0 + \frac{2\alpha_\text{eff}}{\sqrt{3}\gamma'}f$~\cite{mewes2015relaxation, lee2008spin}. Linewidth vs.\ frequency data is shown in Fig.~\ref{FIG6_FMR}(b) for 330 nm thick NFO films on MGO and CGO substrates with Pd top layer. The estimated value of the $\alpha_\text{{eff}}$ of the NFO/MGO and NFO/CGO thin films without Pd top layer are determined to be (22 $\pm$ 0.9) $\times 10^{-4}$ and (1.3 $\pm$ 0.9) $\times 10^{-4}$, respectively (see Fig.~\textcolor{HanRed}{S9} in \textcolor{HanRed}{SI}). After Pd deposition we find an increase in the damping constant and the effective Gilbert damping parameter. The values derived from the fitting of the data in Fig.~\ref{FIG6_FMR}(b) are (2.9 $\pm$ 0.1) $\times 10^{-3}$ and (2.3 $\pm$ 0.1) $\times 10^{-3}$ for the Pd/NFO/MGO and Pd/NFO/CGO films, respectively. The difference in the Gilbert damping parameter of the two films capped with and without Pd can be directly related to the spin current density in the two films which can explain the significant differences in the SSE voltage for the two films~\cite{Chang2017a}. It should be noted that the value of $\alpha_{\text{eff}}$ for NFO/CGO ( (1.3 $\pm$ 0.9) $\times 10^{-4}$) is comparable to the best reported value of YIG/GGG thin films ($\sim$7.35$\times$10$^{-5}$)~\cite{hauser2016yttrium}, suggesting that NFO/CGO is a promising candidate for spin caloritronics and spin transport in general.

\section{IV. Conclusions}

In summary, thin films of NFO exhibit improved structural, interfacial and dynamical properties when grown on lattice-matched substrates. The results clearly show that higher LSSE signal is obtained for the most closely lattice-matched substrates (MGO, CGO). We find that the thinner films on the CGO substrate provide larger LSSE voltage signal as compared to the other heterostructures and this is consistent with the lower value of the effective Gilbert damping of these films. COMSOL Multiphysics$^\circledR$ simulation indicates that the temperature gradient across the film is in the range of tens of mK/\si{\um}. The measurements using the heat flux method also affirm the importance of lattice matching to enhance spin generated voltage signal that also correlates with the FMR results. Apart from this, LSSE measurements provide a qualitative method to study in-plane magnetic anisotropies by varying the angle between the external magnetic field and the direction of the contacts for the detection of the ISHE voltage. Improved quality NFO thin films exhibit damping parameter comparable to that of YIG/GGG, which makes them attractive for spintronics as well as microwave applications. Further improvement of the LSSE efficiency of NFO could be reached by choosing substrates with even less lattice mismatch compared to MGO and CGO.

\section{Acknowledgments}

The work at The University of Alabama was supported by NSF ECCS Grant No.~1509875 and NSF CAREER Award No.~0952929. The work at ORNL (AYB) was supported by the Materials Science and Engineering Division of the Office of Science of the US DOE. RM was supported by a startup funding from Washington University. JG was supported by the Ram\'on y Cajal program (RyC-2012-11709). The Bielefeld group (TP, BP, DM, GR, TK) gratefully acknowledge financial support by the Deutsche Forschungsgemeinschaft (DFG) within the priority program Spin Caloric Transport (SPP 1538).

\bibliographystyle{apsrev4-2}
\bibliography{PRLref}

\end{document}